\documentclass[aps,prc,preprint,amsmath,showpacs,superscriptaddress,nofootinbib]{revtex4}
\usepackage{graphicx,epsfig}
\newcommand{\beqy}{\begin{eqnarray}}
\newcommand{\eeqy}{\end{eqnarray}}
\newcommand{\bmlet}{\begin{subequations}}
\newcommand{\emlet}{\end{subequations}}
\newcommand{\bfdel}{\mbox{\boldmath$\nabla$}}

\begin{document}

\textwidth 16.2 cm
\oddsidemargin -.54 cm
\evensidemargin -.54 cm

\def\gsimeq{\,\,\raise0.14em\hbox{$>$}\kern-0.76em\lower0.28em\hbox  
{$\sim$}\,\,}  
\def\lsimeq{\,\,\raise0.14em\hbox{$<$}\kern-0.76em\lower0.28em\hbox  
{$\sim$}\,\,}

\title{Further explorations of Skyrme-Hartree-Fock-Bogoliubov mass formulas.
XI:  Stabilizing neutron stars against a ferromagnetic collapse.}
\author{N.~Chamel}
\affiliation{Institut d'Astronomie et d'Astrophysique, CP-226, Universit\'e 
Libre de Bruxelles, 1050 Brussels, Belgium}
\author{S.~Goriely}
\affiliation{Institut d'Astronomie et d'Astrophysique, CP-226,
Universit\'e Libre de Bruxelles, 1050 Brussels, Belgium}
\author{J.~M.~Pearson}
\affiliation{D\'ept. de Physique, Universit\'e de Montr\'eal, Montr\'eal 
(Qu\'ebec), H3C 3J7 Canada}
\date{\today}

\begin{abstract}
We construct a new Hartree-Fock-Bogoliubov (HFB) mass model, labeled HFB-18, 
with a generalized Skyrme force. The additional terms that we 
have introduced into the force are density-dependent generalizations of the 
usual $t_1$ and $t_2$ terms, and are chosen in such a way as to avoid the 
high-density ferromagnetic instability of neutron stars that is a general
feature of conventional Skyrme forces, and in particular of the Skyrme forces  
underlying all the HFB mass models that we have developed in the past. The
remaining parameters of the model are then fitted to
essentially all the available mass data, an rms deviation $\sigma$ of 0.585 MeV
being obtained. The new model thus gives almost as good a mass fit as our 
best-fit model HFB-17 ($\sigma$ = 0.581 MeV), and has the 
advantage of avoiding the ferromagnetic collapse of neutron stars.
\end{abstract}

\pacs{21.10.Dr, 21.30.-x, 21.60.Jz, 26.60.Dd, 26.60.Kp}

\maketitle

\section{Introduction}

Astrophysical considerations require that one have available nuclear-mass
models as rigorously based as possible. In this way one might hope
to be able to extrapolate from the mass data, which cluster fairly closely
to the stability line, out towards the neutron drip line, and make reliable
estimates of the masses of nuclei so neutron rich that there is no
hope of measuring  them in the foreseeable future; such nuclei are found in
the outer crusts of neutron stars, and also play a vital role in the r-process 
of nucleosynthesis. To this end we have developed a series of nuclear-mass 
models based on the Hartree-Fock-Bogoliubov (HFB) method with Skyrme and 
contact-pairing forces, together with phenomenological Wigner terms and 
correction terms for the spurious collective energy; all the model parameters 
are fitted to essentially all the experimental mass data (see Ref.~\cite{gcp09}
and references quoted therein). To make the extrapolations to neutron-rich
nuclei as reliable as possible, the underlying Skyrme force in model HFB-9 
\cite{sg05} and all later models was constrained to fit the zero-temperature 
equation of state (EoS) of neutron matter (NeuM), as calculated by Friedman and
Pandharipande~\cite{fp81} (FP) for realistic two- and three-nucleon forces.
(We have so far been unable to obtain mass fits as good as our published ones
when constraining to the more complete, and slightly stiffer, realistic 
neutron-matter EoS labeled A18 + $\delta\,v$ + UIX$^*$~\cite{apr98}. 
Our findings are consistent with a recent analysis of the $\pi^-/\pi^+$ ratio 
in central heavy-ion collisions indicating that this EoS is too 
stiff~\cite{xiao09}.)

Because of the neutron-matter constraint our models can be used to extrapolate 
beyond the neutron drip line and calculate with some confidence the EoS of the 
inner crust of neutron stars~\cite{onsi08} (throughout this paper we assume 
zero temperature). Since the good agreement of our forces with the FP
calculation \cite{fp81} of neutron matter extends to the highest density 
encountered in neutron stars it might be thought that our inner-crust EoS is 
continuous with that of the homogeneous core, the transition between the two
regions taking place at around $0.5\rho_0$, where $\rho_0$ ($\simeq$ 
0.16 nucleons.fm$^{-3}$) is the equilibrium density of symmetric nuclear matter
(SNM). However, in fitting our forces
to the neutron matter of FP \cite{fp81}, we {\it assume} that our ground state
is spin unpolarized, but in fact the underlying Skyrme forces of all our 
models, like all conventional Skyrme forces of the form (\ref{1}), predict that
beyond a certain supernuclear density the ground state of NeuM becomes 
ferromagnetic, i.e., at least partially polarized~\cite{vida84,kw94,mar02}. In 
the case of the Skyrme force BSk17, the force underlying our best-fit 
model, HFB-17~\cite{gcp09}, complete polarization sets in at a density of 
$\rho_{frmg} = 1.24\,\rho_0$. On the other hand, microscopic calculations 
using different realistic forces and different 
methods~\cite{fan01, bomb06, sam07, bord08,vid02} all predict no such 
polarization, at least up to about $5\rho_0$. Moreover, in the case of all our 
own previously published HFB forces the
predicted ferromagnetic state is unstable against collapse, the energy becoming
more and more negative as the density increases (see, for example, the lowest 
curve in Fig.~\ref{fig1}, constructed for force BSk17). This predicted 
ferromagnetic collapse sets in at densities that are certainly 
encountered in the cores of all neutron stars, and is contradicted by the very 
existence of neutron stars.

Actually, the core of neutron stars does not consist of pure NeuM but rather of
so-called neutron-star matter (N*M), which just below the crust is composed of 
neutrons with a weak admixture of proton-electron pairs in beta
equilibrium (for simplicity we will assume that these are the only particles present 
also at higher densities). 
But if NeuM were indeed unstable against collapse then N*M would 
be likewise, since, at the very least, it would always be energetically 
advantageous for N*M to spontaneously transform into NeuM through electron 
capture beyond some critical density. Thus the stability of NeuM against 
collapse is a necessary condition for the existence of neutron stars. 
But even if no such instability is implied, there is 
a general tendency for Skyrme forces of the conventional form (\ref{1}) to
lead to the ground state of N*M being polarized. However, calculations based on
realistic forces show the ground state of N*M, like that of NeuM,
to be unpolarized at all densities~\cite{bord08b}.

Our purpose here is to show that by adding suitable terms to the Skyrme 
force it is possible to 
eliminate the anomalous prediction of a ferromagnetic transition in neutron 
stars, with essentially no impact on the high-quality fits to the mass data
that we have previously obtained with conventional Skyrme forces.
(An alternative approach to this problem has been followed by 
Margueron {\it et al.} \cite{marg09a, marg09}.) 
In Section II we discuss in more detail the nature of the spurious
transition to a ferromagnetic state in NeuM and N*M associated with 
conventional Skyrme forces of the form (\ref{1}), considering not only our own 
BSk17 \cite{gcp09}, but also the widely used SLy4 \cite{cha98}, which was specifically
constructed for neutron-star calculations. Section III
shows how extra terms in the Skyrme force can stop this spurious transition,
while in Section IV we describe the new mass fit. Our conclusions are 
summarized in Section V, and in the Appendix we present the full formalism for
the generalized form of Skyrme force used here.  

\section{Ferromagnetic instability}

The Skyrme forces that we have used in all our previously published HFB models 
have the conventional form 
\beqy
\label{1}
v_{i,j} & = & 
t_0(1+x_0 P_\sigma)\delta({\pmb{r}_{ij}})
+\frac{1}{2} t_1(1+x_1 P_\sigma)\frac{1}{\hbar^2}\left[p_{ij}^2\,
\delta({\pmb{r}_{ij}}) +\delta({\pmb{r}_{ij}})\, p_{ij}^2 \right]\nonumber\\
& &+t_2(1+x_2 P_\sigma)\frac{1}{\hbar^2}\pmb{p}_{ij}.\delta(\pmb{r}_{ij})\,
 \pmb{p}_{ij}
+\frac{1}{6}t_3(1+x_3 P_\sigma)\rho(\pmb{r})^\alpha\,\delta(\pmb{r}_{ij})
\nonumber\\
& &+\frac{\rm i}{\hbar^2}W_0(\mbox{\boldmath$\sigma_i+\sigma_j$})\cdot
\pmb{p}_{ij}\times\delta(\pmb{r}_{ij})\,\pmb{p}_{ij}  \quad ,
\eeqy
where $\pmb{r}_{ij} = \pmb{r}_i - \pmb{r}_j$, $\pmb{r} = (\pmb{r}_i + 
\pmb{r}_j)/2$, $\pmb{p}_{ij} = - {\rm i}\hbar(\pmb{\nabla}_i-\pmb{\nabla}_j)/2$
is the relative momentum, $P_\sigma$ is the two-body spin-exchange 
operator, and $\rho(\pmb{r}) = \rho_n(\pmb{r}) + \rho_p(\pmb{r})$ is the total 
local density, $\rho_n(\pmb{r})$ and $\rho_p(\pmb{r})$ being the neutron and 
proton densities, respectively. 

\begin{figure}
\centerline{\epsfig{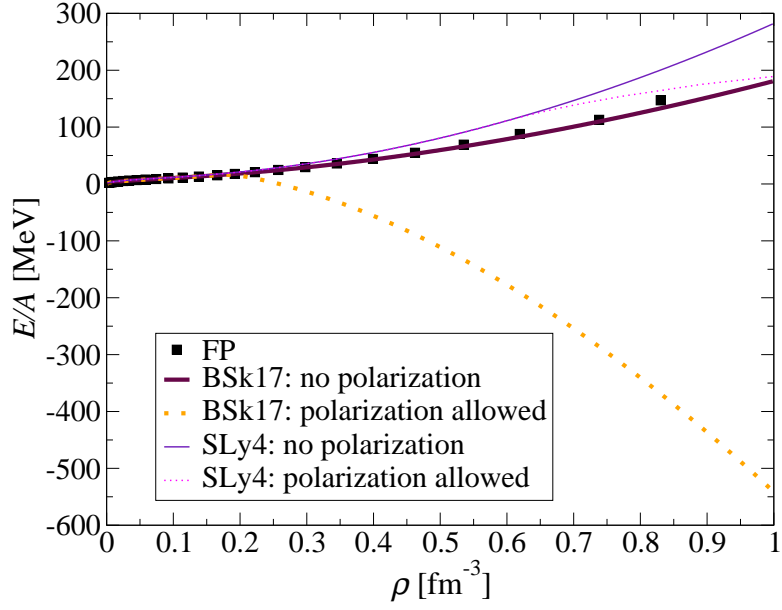}}
\caption{(Color online) Energy per nucleon ($T$= 0) for pure neutron matter (NeuM) with forces
BSk17 and SLy4. The squares are the results 
of the realistic calculations from Friedman and Pandharipande~\cite{fp81}.}
\label{fig1}
\end{figure}

\begin{figure}
\centerline{\epsfig{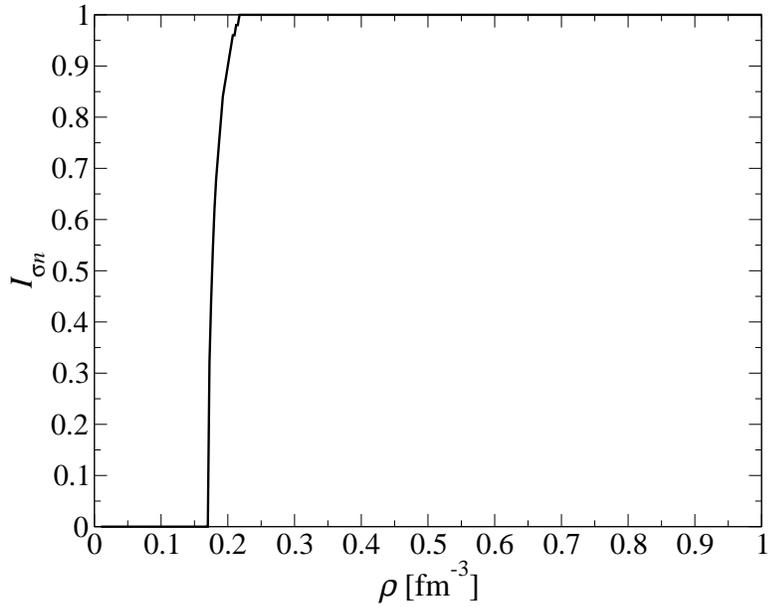}}
\caption{Neutron polarization $I_{\sigma n}$ for pure neutron matter (NeuM)
with force BSk17.} 
\label{fig2}
\end{figure}

The fit of the parameters of this form of force to the nuclear masses, and the 
various other constraints mentioned above, leave us absolutely no freedom to
avoid an unphysical ferromagnetic collapse of NeuM. The situation for our force
BSk17 is shown in Figs.~\ref{fig1} and~\ref{fig2}. (The corresponding curves
for force SLy4 \cite{cha98} shown in these two figures are discussed below.) 
These curves have been calculated using Eq. (C.~14) of 
Bender {\it et al.}~\cite{bend02}, which gives the energy per nucleon of
homogeneous nuclear matter with arbitrary charge asymmetry and degree of
polarization; we define the latter quantity for nucleons of charge type $q$ 
($q = n$ or $p$) by 
\beqy\label{2}
I_{\sigma q} = \frac{\rho_{q \uparrow} - \rho_{q \downarrow}}
{\rho_{q \uparrow} + \rho_{q \downarrow}} \quad .
\eeqy
The curve labeled ``BSk17: no polarization'' in Fig.~\ref{fig1} shows the
energy per neutron calculated under the constraint $I_{\sigma n} = 0$, while
the curve ``BSk17: polarization allowed'' shows the same quantity calculated
at each density by minimizing the energy per neutron with respect to 
$I_{\sigma n}$. The first of these curves is seen to be in excellent agreement
with the realistic EoS of NeuM given by FP~\cite{fp81}, but the second curve
shows that this agreement is destroyed when polarization is allowed, the system
collapsing. The corresponding value of $I_{\sigma n}$ for BSk17 is shown in 
Fig.~\ref{fig2}; the rapidity with which polarization sets in with increasing
density will be seen. 

\begin{figure}
\centerline{\epsfig{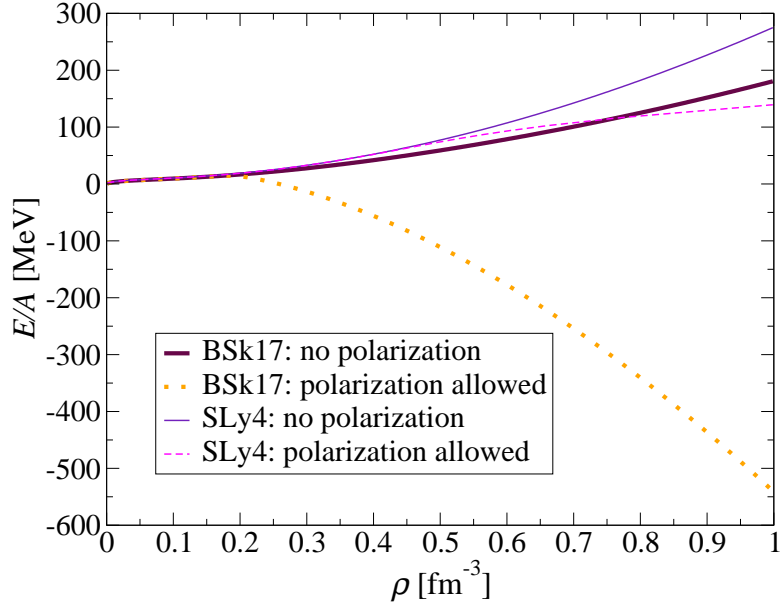}}
\caption{(Color online) Energy per nucleon ($T$= 0) for neutron-star matter (N*M) with forces
BSk17 and SLy4.}
\label{fig3}
\end{figure}

\begin{figure}
\centerline{\epsfig{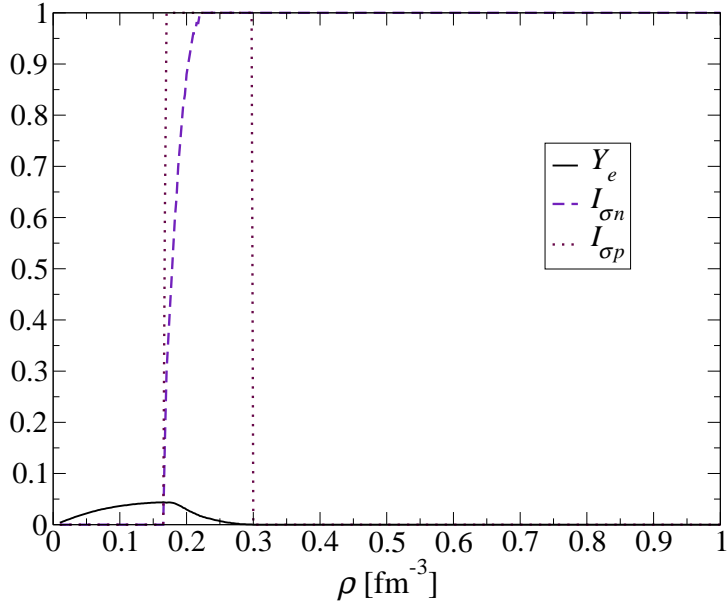}}
\caption{(Color online) Proton fraction $Y_e$, neutron and proton polarizations 
$I_{\sigma n}$ and $I_{\sigma p}$, respectively, for neutron-star 
matter (N*M) with force BSk17.}
\label{fig4}
\end{figure}

As for N*M, to the nuclear energy corresponding to the
Skyrme force in this neutron-proton mixture we have to add the electron kinetic
energy, for the density of which we take the
exact expression (see, for example, Section 24.6c of Ref.~\cite{cg})
\beqy\label{3}
u_e = \frac{mc^2}{24\pi^2}\left(\frac{mc}{\hbar}\right)^3\left\{-8x^3 +
3x(1 + 2x^2)(1 + x^2)^{1/2} - 3\sinh^{-1}x\right\} \quad ,
\eeqy
where
\beqy\label{4}
x = \frac{\hbar}{mc}\left(3\pi^2\rho_e\right)^{1/3},
\eeqy
$\rho_e$ being the electron density (equal to the proton density $\rho_p$,
because of electric-charge neutrality). We  minimize the total energy
per nucleon with respect to the proton fraction $Y_e = \rho_p/\rho$,
while imposing the constraint $I_{\sigma n} =  I_{\sigma p} = 0$, and obtain
the curve labeled ``BSk17: no polarization'' in Fig.~\ref{fig3}. If we next
minimize the total energy with respect to $Y_e, I_{\sigma n}$ and
$I_{\sigma p}$ we obtain the curve labeled ``BSk17: polarization allowed'' in
Fig.~\ref{fig3}; the corresponding values of  $Y_e, I_{\sigma n}$ and
$I_{\sigma p}$ as a function of density are shown in Fig.~\ref{fig4}. It will
be seen that once polarization is admitted BSk17 leads in N*M to the same 
instability with respect to collapse that we found in NeuM (Fig.~\ref{fig1}). 

{\it The case of Skyrme force SLy4.} Figs.~\ref{fig1} and~\ref{fig3} also show 
the corresponding curves for force SLy4~\cite{cha98} in NeuM and N*M, 
respectively. It will be seen that in NeuM (Fig.~\ref{fig1}), the onset of 
polarization has been postponed to $\rho_{frmg} = 4.4\rho_0$, and that although
there is a considerable softening of the EoS there is no collapse, at least at
neutron-star densities. However, Fig.~\ref{fig3} shows that although
there is still no collapse in N*M, the transition to a spurious ferromagnetic 
state, and the associated softening of the EoS, takes place at the much lower 
density of $\rho_{frmg} = 2.5\rho_0$, which certainly lies within the range of 
densities found in neutron stars. This shows that stability against a 
ferromagnetic transition in NeuM at a given density does not guarantee
stability in N*M at the same density. 
(A similar conclusion has been reached in Ref.~\cite{marg09a}.)

The fact that SLy4 has greater stability against polarization than does BSk17
is a result of setting $x_2 = -1$. Now within the framework of a 
conventional Skyrme force of the form~(\ref{1}) we have been unable to find an 
acceptable mass fit for $x_2 = -1$, and indeed SLy4 performs badly
as a global mass model, the rms deviation for the even-even nuclei being quoted
as 5.1 MeV \cite{sto03}. This could have implications for the composition of 
both the outer and inner crusts, and might explain the differences between 
SLy4 and BSk14 predictions shown in Figures 1 and 2 of Ref.\cite{onsi08}.
It seems that it is impossible to have both a 
good mass fit and stability against polarization with the conventional form
(\ref{1}) of Skyrme force.  

\section{Stability restored}

We note now that the form~(\ref{1}) does not exhaust the possibilities for a 
Skyrme-type force, and we shall consider here two extra terms, writing our 
complete Skyrme force as
\beqy
\label{5}
v^{\prime}_{i,j} &=& v_{i,j} +
\frac{1}{2}\,t_4(1+x_4 P_\sigma)\frac{1}{\hbar^2} \left\{p_{ij}^2\,
\rho({\pmb{r}})^\beta\,\delta({\pmb{r}}_{ij}) +
\delta({\pmb{r}}_{ij})\,\rho({\pmb{r}})^\beta\, p_{ij}^2 \right\} \nonumber\\
&+&t_5(1+x_5 P_\sigma)\frac{1}{\hbar^2}{\pmb{p}}_{ij}.
\rho({\pmb{r}})^\gamma\,\delta({\pmb{r}}_{ij})\, {\pmb{p}}_{ij}
 \quad .
\eeqy
The $t_4$ and $t_5$ terms are density-dependent generalizations of the $t_1$ 
and $t_2$ terms, respectively. 
The formalism for this generalized Skyrme
force is developed in Appendix~\ref{formal}, where we show in particular that
Eq.~(C.~14) of Bender {\it et al.}~\cite{bend02} for homogeneous nuclear 
matter of arbitrary charge asymmetry and polarization can be generalized to
include the new terms simply by making the substitutions of Eqs.~(\ref{D3a}) -
(\ref{D3d}).

We consider now just the simpler case of NeuM, the stabilization of which is a 
necessary condition. For complete polarization in the presence of the 
conventional Skyrme force~(\ref{1}) the energy per nucleon is given by 
\cite{kw94}
\beqy\label{7}
e = \frac{3\hbar^2}{10M_n}(6\pi^2\rho)^{2/3} +  
\frac{3}{10}(6\pi^2)^{2/3}t_2(1 + x_2)\rho^{5/3} \quad .
\eeqy
The only Skyrme term operative here is the one in $t_2$. 
The ferromagnetic collapse of neutron matter predicted by all our Skyrme forces 
at supernuclear densities arises from the fact that the combination
$t_2(1 + x_2)$ is always negative. However, stability can always be enforced 
by adding a new repulsive term in $t_5$, as can be seen from Eqs.~(\ref{D3c}) 
and (\ref{D3d}). But we do not want to disturb in any way the unpolarized 
configuration of NeuM, since we know from the experience with our recent mass 
models that with the conventional form of Skyrme force (\ref{1}) alone it is
easy to fit the realistic EoS of FP~\cite{fp81}, and wish to do so here with 
the new model. As the energy per nucleon of this latter configuration in the 
presence of the conventional Skyrme force~(\ref{1}) is given by 
\beqy\label{8}
e &=& \frac{3\hbar^2}{10M_n}(3\pi^2\rho)^{2/3} + \frac{1}{4}t_0(1 - x_0)\rho
+ \frac{1}{24}t_3(1 - x_3)\rho^{\alpha + 1}  \nonumber \\
&+& \frac{3}{40}(3\pi^2)^{2/3}\left\{t_1(1 - x_1) + 3t_2(1 + x_2)\right\}
\rho^{5/3}   
\eeqy
(see Eq.~(\ref{C13})), it follows from Eqs. (\ref{D3a}) and (\ref{D3b}) that 
the $t_5$ term can be completely canceled in unpolarized NeuM by adding a $t_4$
term with its parameters constrained by 
\beqy\label{9}
\beta = \gamma
\eeqy
and
\beqy\label{10}
t_4(1 -x_4) = -3t_5(1 + x_5) \quad .
\eeqy

It is now highly convenient to require that the stabilizing terms in $t_4$ and 
$t_5$ cancel completely in unpolarized nuclear matter of {\it any} charge 
asymmetry. This leads, using Eq.~(\ref{C1}), to a second condition on $t_4$, 
\beqy\label{11}
t_4 = -\frac{1}{3}t_5(5 + 4x_5)  \quad ,
\eeqy
which, combined with Eq.~(\ref{10}), leads to
\beqy\label{12}
x_4 = -\frac{4 + 5x_5}{5 + 4x_5} \quad .
\eeqy
Thus all three parameters of the $t_4$ term will be completely determined by  
the parameters of the $t_5$ term, i.e., $t_5, x_5$ and $\gamma$. These
latter three parameters leave us with ample flexibility for stabilizing NeuM 
against polarization, and possible collapse. We stress, however, that it will
also be necessary to check the stability of N*M. 

\section{The HFB-18 mass model}

Even though the new terms exactly cancel in homogeneous nuclear matter they
will not do so in finite nuclei, and we cannot simply add them on to the
BSk17 force. Rather it will be necessary to make a complete refit of all the
model parameters to the mass data.

Our HFB calculations for finite nuclei are performed exactly as 
for the HFB-17 model \cite{gcp09}. In particular, the treatment of pairing,
which is neglected in the neutron-matter constraints discussed in the previous
section, is highly realistic. As usual, we take a contact pairing force that 
acts only between nucleons of the same charge state $q$, 
\beqy
\label{13}
v^{\rm pair}_q(\pmb{r_i}, \pmb{r_j})= 
v^{\pi\,q}[\rho_n(\pmb{r}),\rho_p(\pmb{r})]~\delta(\pmb{r}_{ij})\quad ,
\eeqy
where the strength $v^{\pi\,q}[\rho_n,\rho_p]$ is a functional of both the 
neutron and proton densities. But instead of postulating a simple functional 
form for the density dependence, as is usually done, we construct the pairing 
force by solving the HFB equations in uniform asymmetric nuclear matter with
the appropriate neutron and proton densities, requiring that the 
resulting gap reproduce exactly, as a function of density, the microscopic 
$^1S_0$ pairing gap calculated with realistic forces~\cite{cao06}.
We follow our usual practice of allowing the proton pairing strength to be 
different from the neutron pairing strength, and for allowing each of these
strengths to depend on whether there is an even or odd number of nucleons of
the charge type in question. These extra degrees of freedom are taken into
account by multiplying the value of $v^{\pi\,q}[\rho_n, \rho_p]$, as
determined by the nuclear-matter calculations that we have just described,
with renormalizing factors $f^{\pm}_q$, where $f^+_p, f^-_p$ and $f^-_n$ are 
free, density-independent parameters to be included in the mass fit, and  
we set $f^+_n = 1$. 

To the HFB energy calculated for the Skyrme and pairing forces we add 
a Wigner correction, 
\beqy\label{14}
E_W = V_W\exp\Bigg\{-\lambda\Bigg(\frac{N-Z}{A}\Bigg)^2\Bigg\}
+V_W^{\prime}|N-Z|\exp\Bigg\{-\Bigg(\frac{A}{A_0}\Bigg)^2\Bigg\} \quad ,
\eeqy
which contributes significantly only for light nuclei with $N$ close to $Z$. 
Our treatment of this correction is purely phenomenological, although physical 
interpretations of each of the two terms can be made \cite{sg02,cgp08}.

A second correction that must be made is to subtract from the HFB energy
an estimate for the spurious collective energy. As described in 
Ref.~\cite{cgp08}, the form we adopt here is
\beqy\label{15}
E_{coll}= E_{rot}^{crank}\Big\{b~\tanh(c|\beta_2|) +
d|\beta_2|~\exp\{-l(|\beta_2| - \beta_2^0)^2\}\Big\} \quad  ,
\eeqy
in which $E_{rot}^{crank}$ denotes the cranking-model value of the rotational
correction and $\beta_2$ the quadrupole deformation, while all other parameters
are free fitting parameters.

The final correction that we make is to drop Coulomb exchange. This is a device
that we have successfully adopted in our most recent models,
beginning with HFB-15 \cite{gp08},  and it can be interpreted as
simulating neglected effects such as Coulomb correlations, charge-symmetry 
breaking of the nuclear forces, and vacuum polarization. 
 
The parameters of the model, i.e., of the Skyrme and pairing forces, and of the
Wigner and collective corrections, are fitted to the same set of mass data as 
are all our models since HFB-9~\cite{sg05}, namely, the 2149 measured masses of
nuclei with $N$ and $Z \ge$ 8 given in the 2003 Atomic Mass 
Evaluation~\cite{audi03}. This fit is subject to the constraints on both 
unpolarized and polarized NeuM discussed in the previous section, as well as 
our usual requirement that the isoscalar effective mass $M_s^*$ take the 
realistic value of 0.8$M$ in SNM at the equilibrium density $\rho_0$. The 
values of the Skyrme, pairing and Wigner parameters resulting from this fit 
are shown in Table~\ref{tab1} ($\varepsilon_{\Lambda}$ is the pairing cutoff
parameter \cite{cgp08,gcp09}): this defines the BSk18 ``force". The parameters
of the collective correction are shown in Table~\ref{tab2}. 

\begin{figure}
\centerline{\epsfig{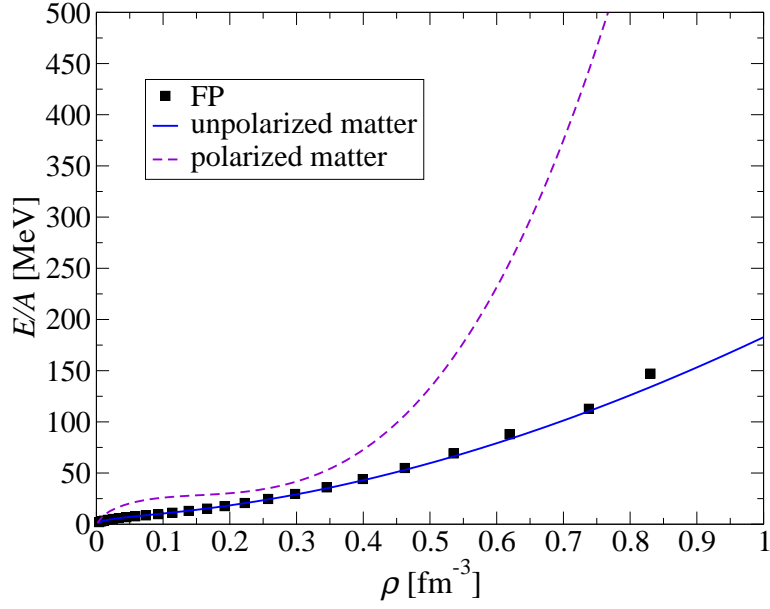}}
\caption{(Color online) Energy per nucleon ($T$=0) for pure neutron matter (NeuM),
polarized and unpolarized, with force BSk18. The squares show the results of
the realistic calculations of Friedman and Pandharipande~\cite{fp81}.}
\label{fig5}
\end{figure}

\begin{figure}
\centerline{\epsfig{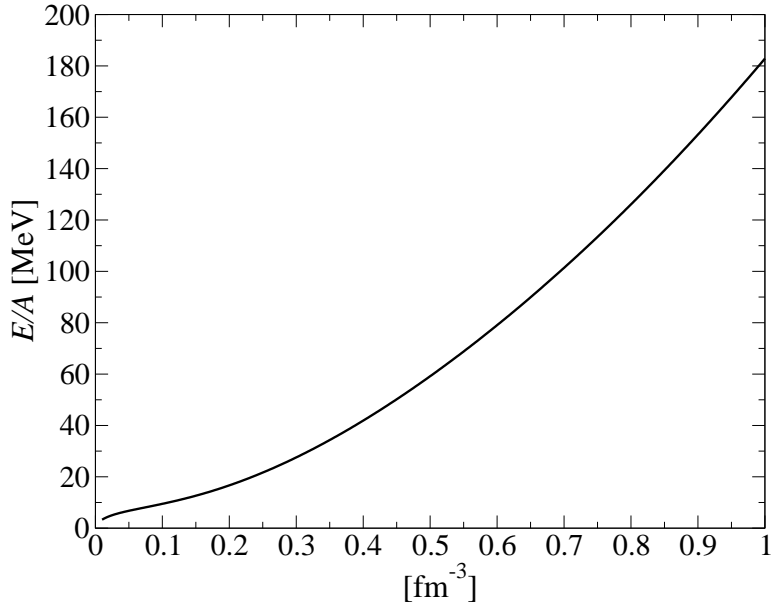}}
\caption{Energy per nucleon ($T$= 0) for neutron-star matter (N*M) with force BSk18. 
System stable against polarization in ground state.}
\label{fig6}
\end{figure}

\begin{figure}
\centerline{\epsfig{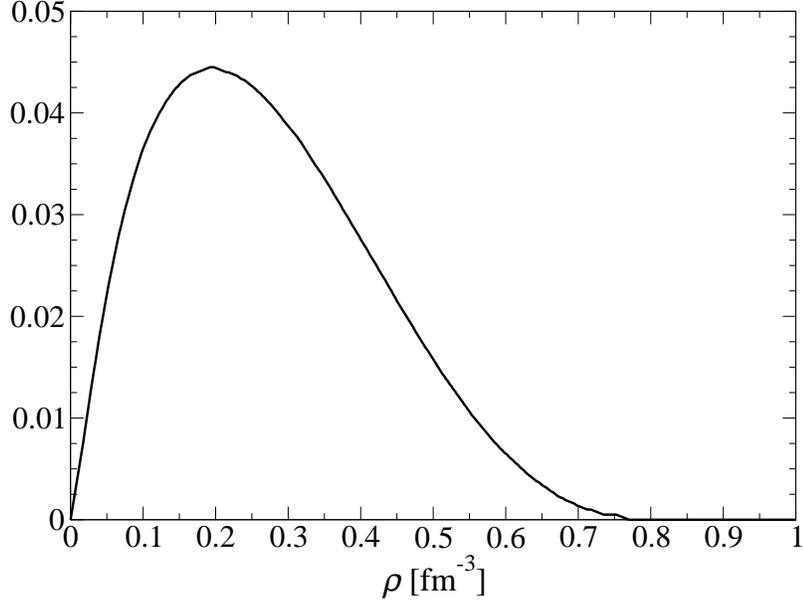}}
\caption{Proton fraction $Y_e$ for neutron-star matter (N*M) with force BSk18.}
\label{fig7}
\end{figure}

The starting point for the new fit was the force BSk17 \cite{gcp09}, and we 
stress that the parameter search was far from exhaustive, 
particularly with regards to $t_5, x_5$ and $\gamma$. 
Our choice of these latter parameters was
somewhat arbitrary, being limited only by the requirements that the 
energy-density curve for polarized NeuM lie entirely above that for 
unpolarized NeuM, and that there be no significant deterioration in the quality
of the overall mass fit. That the first of these requirements is satisfied is 
seen in Fig.~\ref{fig5}, which automatically guarantees the stability of NeuM
against polarization. Proceeding as in Section II, we have checked also that 
N*M is stable against polarization over the same density range, as seen in 
Figs.~\ref{fig6} and~\ref{fig7} (force BSk17 gives virtually identical results 
in N*M, provided we do not allow the ground state to be polarized for this 
force). Likewise, it is apparent in the first line of Table~\ref{tab3} that we 
have satisfied the requirement on the quality of the mass fit, and indeed it 
can be seen from this table that in some respects the new model is better than 
the old. The fact that
large regions of the extended parameter space remain unexplored allows ample
scope for future improvement, both with respect to the quality of the mass fit
and conformity of the force to reality. Note, however, that as long
as our force is constrained by Eqs. (\ref{9}), (\ref{11}) and (\ref{12}) the 
$t_4$ and $t_5$ components of the force will not contribute to the effective
mass, and it will not be possible to exploit the advantages of a surface-peaked
effective mass \cite{fpt01}: see Eq.~(\ref{B5}).

The comparison of the new parameter set with the original BSk17 set in 
Table~\ref{tab1} shows that the changes in the $t_1$ and $t_2$ terms are much 
greater than those in the $t_0$ and $t_3$ terms; this simply 
reflects the fact that the new $t_4$ and $t_5$ terms cancel in homogeneous 
unpolarized matter, and act only through the gradient terms to which they 
give rise. (Their contribution to the binding is nevertheless far from 
negligible, amounting to 38 MeV in the case of $^{208}$Pb.) For 
the same reason there is very little difference between the 
BSk17 and BSk18 values of the droplet-model parameters \cite{ms69} shown in
the first five lines of Table~\ref{tab4}. 
There is likewise very little change
in most of the Landau parameters shown in the last eight lines of 
Table~\ref{tab4}. Only $G_0$ and $G_1$ show any substantial differences
between BSk17 and BSk18, these  being the only two Landau parameters to which 
the $t_4 $ and $t_5$ components can contribute when the condition~(\ref{9}) 
holds and the particular choice $t_4=t_5$ is made. In particular, the 
generalized Skyrme force BSk18 leads to $G_0$ being larger than -1 in SNM for 
all densities, thus removing the spin instability that was predicted by our 
previous force BSk17, as can be seen in Figure~\ref{fig9}. On the other hand,
a spin-isospin instability still occurs in SNM at about the same high density 
as that found for BSk17, the two forces giving very similar values of 
$G_0^\prime$ for all densities (see Figure~\ref{fig10}). However, 
insofar as beta-equilibrated N*M has, as we have seen, a large
neutron excess, this spin-isospin instability in SNM is of no consequence
for neutron stars.

\begin{figure}
\centerline{\epsfig{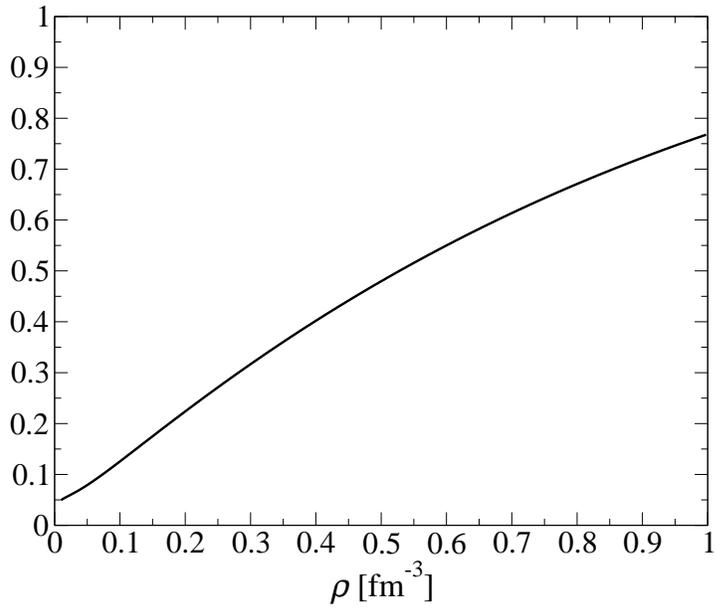}}
\caption{Speed of sound $v_s$ (in units of the light speed $c$) for pure 
neutron-matter (NeuM) with force BSk18.}
\label{fig8}
\end{figure}

\begin{figure}
\centerline{\epsfig{figure=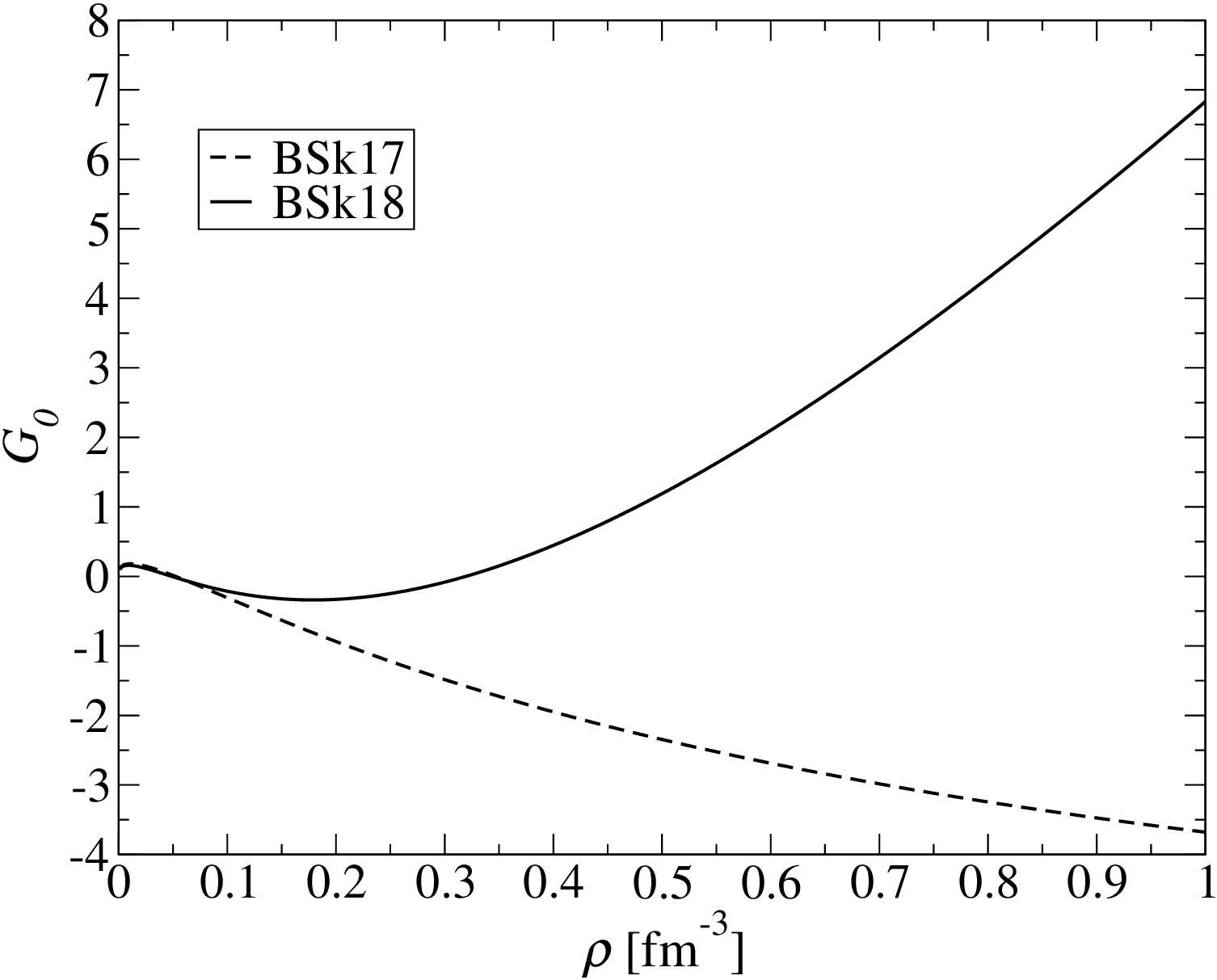,height=8.0cm}}
\caption{Landau parameter $G_0$ in symmetric nuclear matter for forces BSk17 (dashed line) and 
BSk18 (solid line).}
\label{fig9}
\end{figure}

\begin{figure}
\centerline{\epsfig{figure=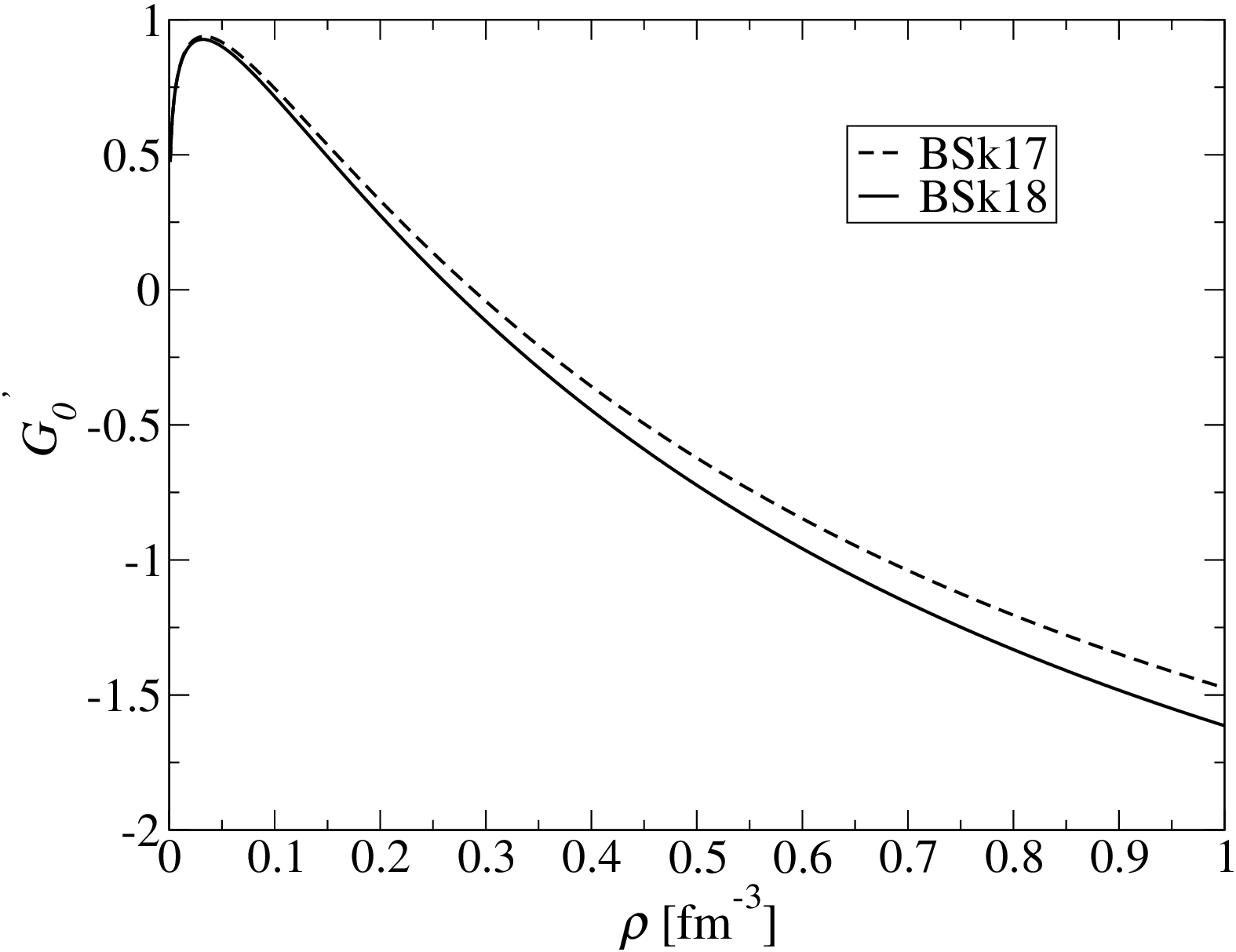,height=8.0cm}}
\caption{Landau parameter $G_0^\prime$ in symmetric nuclear matter for forces BSk17 (dashed line) and 
BSk18 (solid line).}
\label{fig10}
\end{figure}

Finally, we have checked that with the new interaction, BSk18, causality
is satisfied in NeuM for all densities found in neutron stars, i.e, the
velocity of sound $v_s$ is smaller than the velocity of light $c$. The former
is given by the relativistic expression
\beqy\label{caus1}
v_s = c\sqrt{\frac{\partial\,P(\rho)}{\partial\,\mathcal{E}(\rho)}}
\quad  ,
\eeqy
in which $P(\rho)$ is the pressure,
\beqy\label{caus2}
P(\rho) = \rho^2\frac{\partial\,e(\rho)}{\partial\,\rho}
\eeqy
and $\mathcal{E}(\rho)$ is the total energy density,
\beqy\label{caus3}
\mathcal{E}(\rho) = \rho\left(e(\rho) + M_nc^2\right) \quad ,
\eeqy
$e(\rho)$ being simply the energy per neutron, as given by 
Eq.~(\ref{C13}). The calculated value of $v_s/c$ for BSk18 is shown as a 
function of $\rho$ in Fig.~\ref{fig8}.

Using this interaction BSk18, we have constructed a complete mass table,
labeled HFB-18, running from one drip line to the other over the range $Z$ and
$N \ge$ 8 and $Z \le$ 110. The results are very similar to those for HFB-17, 
the rms difference between all 8389 predictions being 0.433 MeV, and the mean 
difference (HBF-18 - HFB-17) 0.198 MeV. We have also calculated the spins and 
parities of all these nuclei; only for about 10\% of these nuclei is there a
difference with respect to the HFB-17 prediction.

\section{Conclusions.} 

We have extended our earlier Skyrme-HFB mass models by the inclusion of terms 
that are density-dependent generalizations of the usual $t_1$ and $t_2$ terms.
We have shown that these new terms can be chosen in such a way as to prevent 
the high-density ferromagnetic collapse of neutron stars that was a general 
feature of our previous HFB mass models, without compromising the excellent fit
to the mass data that we obtained in the past, and without relaxing any of the 
previously imposed constraints of conformity to reality. The mass predictions 
made by the new model, HFB-18, are, in fact, very similar to those made by the 
preceding model, HFB-17~\cite{gcp09}. These two models not only give better 
fits to the mass data than does any other published model except that of 
Duflo and Zucker~\cite{dz95}, but they are also by far the most microscopically
founded models, and in particular their underlying interactions (BSk17 and
BSk18, respectively) have been fitted to realistic calculations of both the 
EoS and the $^1S_0$ gap of neutron matter. They can thus be expected 
to make more reliable predictions of the highly neutron-rich nuclei that appear
in the outer crust of neutron stars and that are involved in the r-process.
Moreover, these mass models can be used to extrapolate beyond the drip line to 
the inner crust of neutron stars, using the respective interactions to 
calculate the EoS in this region. Our confidence in this extrapolation derives
not only from the fit of the interactions to neutron matter but also from the 
precision fit to masses, which means that the presence of protons and the
existence of inhomogeneities in the inner crust are well represented. 

Finally, pursuing our aim of a unified treatment of the different regions of 
a neutron star, we can use these effective interactions to calculate the EoS of the 
homogeneous core (at least in its outer parts where no complications from the 
possible appearance of hyperons and other particles arise).  
Of course, for pure NeuM nothing new can be obtained in this way
beyond what has already been given by the realistic calculations to which our
effective interactions have been fitted, but these interactions can then be 
reliably used to calculate N*M, which is not treated in the realistic 
calculations of FP~\cite{fp81}. Another important application of these
effective interactions, which would hardly be practical with realistic forces,
is to make a detailed study of the transition between the inner crust and the 
fluid core.

We have seen that for N*M, as for masses, the two interactions BSk17 and BSk18 give very 
similar results, provided we assume that the ground state for the former is 
unpolarized, which in fact is not the case. Here lies the basic difference
between the two interactions: with BSk18 the ground state of NeuM and N*M is 
indeed unpolarized at all densities prevailing in neutron stars.  
This elimination of the spurious polarization is the essential contribution of
this paper, made possible by the introduction of the terms in $t_4$ and $t_5$.
However,
we have so far made only a partial search in the new, extended
parameter space, and there remains the possibility not only of improving the
fit to the mass data still further but also of imposing further realistic
constraints on the Skyrme force.

\appendix

\section{Formalism for generalized Skyrme force.   } 
\label{formal}

The formalism for the conventional Skyrme force (\ref{1}), with expressions for
the energy density, single-particle fields, etc., has been given many times, 
and is conveniently summarized in Brack {\it et al.}~\cite{bgh85}. The 
extension to cover the $t_4$ terms was given by 
Farine {\it et al.}~\cite{fpt01}, but the $t_5$ terms have not, to our 
knowledge, been dealt with before, except for the special case where the 
exponent $\gamma$ is set equal to 1~\cite{kre75}. 

{\it Energy density.}

Assuming invariance under time reversal, the HFB energy is written as the integral of a purely 
local energy-density functional
\beqy\label{A0a}
E_{\rm HFB} = \int \mathcal{E}_{\rm HFB}(\pmb{r})\,{\rm d}^3\pmb{r} \quad ,
\eeqy 
where
\beqy\label{A0b}
\mathcal{E}_{\rm HFB}(\pmb{r}) &=& \mathcal{E}_{\rm Sky}\Big[\rho_n(\pmb{r}),
\bfdel\rho_n(\pmb{r}), \tau_n(\pmb{r}),{\bf J}_n(\pmb{r}), \rho_p(\pmb{r}),
\bfdel\rho_p(\pmb{r}), \tau_p(\pmb{r}),{\bf J}_p(\pmb{r})\Big]  \nonumber \\
 &+&\mathcal{E}_{\rm Coul}\Big[\rho_p(\pmb{r})\Big] +
\mathcal{E}_{\rm pair}\Big[\rho_n(\pmb{r}), \tilde{\rho}_n(\pmb{r}),
\rho_p(\pmb{r}), \tilde{\rho}_p(\pmb{r})\Big] \quad .
\eeqy
The first term here, the energy density for the Skyrme force of this paper, is
given by 
\beqy\label{A1}
\mathcal{E}_{\rm Sky}&=& \sum_{q=n,p}\frac{\hbar^2}{2M_q}\tau_q +
\frac{1}{2} t_0 \Biggl[ \left(1+\frac{1}{2} x_0\right) \rho^2 -
\left(\frac{1}{2} + x_0\right) \sum_{q=n,p} \rho_q^2 \Biggr]  \nonumber \\
&+& \frac{1}{4} t_1 \Biggl[ \left(1+\frac{1}{2} x_1\right) \left(\rho \tau + 
\frac{3}{4} (\nabla \rho)^2\right) - \left(\frac{1}{2} + x_1\right) 
\sum_{q=n,p} \left(\rho_q\tau_q + \frac{3}{4} (\nabla \rho_q)^2\right)\Biggr]  
\nonumber \\
&+& \frac{1}{4} t_2 \Biggl[ \left(1+\frac{1}{2} x_2\right)  \left(\rho\tau 
- \frac{1}{4} (\nabla \rho)^2\right) + \left(\frac{1}{2} + x_2\right) 
\sum_{q=n,p} \left(\rho_q\tau_q - \frac{1}{4} (\nabla \rho_q)^2\right)\Biggr]  
\nonumber \\
&+&  \frac{1}{12}t_3 \rho^\alpha \Biggl[ \left(1+\frac{1}{2} x_3\right)\rho^{2}
 - \left(\frac{1}{2} + x_3\right)\sum_{q=n,p} \rho_q^{2} \Biggr] \nonumber \\
&+& \frac{1}{4} t_4 \Biggl[ \left(1+\frac{1}{2} x_4\right)   \left(\rho \tau + 
\frac{3}{4} (\nabla \rho)^2\right) - \left(\frac{1}{2} + x_4\right) 
\sum_{q=n,p} \left(\rho_q\tau_q + \frac{3}{4} (\nabla \rho_q)^2\right)\Biggr] 
\rho^\beta \nonumber \\
&+& \frac{\beta}{8} t_4 \Biggl[ \left(1+\frac{1}{2} x_4\right) \rho 
(\pmb{\nabla}\rho)^2 - \left(\frac{1}{2}+ x_4\right)\pmb{\nabla}\rho\cdot
\sum_{q=n,p} \rho_q \pmb{\nabla}\rho_q\Biggr]\rho^{\beta - 1} \nonumber \\
&+& \frac{1}{4} t_5 \Biggl[ \left(1+\frac{1}{2} x_5\right)  \left(\rho\tau - 
\frac{1}{4} (\nabla \rho)^2\right) + \left(\frac{1}{2} + x_5\right) 
\sum_{q=n,p} \left(\rho_q\tau_q - \frac{1}{4} (\nabla \rho_q)^2\right)\Biggr] 
\rho^\gamma \nonumber \\
&-& \frac{1}{16} \left(t_1 x_1+t_2 x_2\right) J^2 + \frac{1}{16} \left(t_1 - t_2 \right) \sum_{q=n,p} J_q^2  \nonumber \\
&-& \frac{1}{16} \left(t_4 x_4\rho^\beta+t_5 x_5\rho^\gamma\right)J^2 + 
\frac{1}{16} \left(t_4\rho^\beta - t_5\rho^\gamma \right)\sum_{q=n,p} J_q^2  
\nonumber \\
&+& \frac{1}{2} W_0\left( \pmb{J}\cdot \pmb{\nabla}\rho + \sum_{q=n,p} \pmb{J_q}\cdot \pmb{\nabla}\rho_q \right) \, .
\eeqy
It will be seen that the $t_5$ term has the same form as the $t_2$ term, 
multiplied by the $\rho^\gamma$ factor. No such simple relation between the 
$t_4$ and $t_1$ terms is apparent, because we have eliminated terms containing
a Laplacian, through integration by parts over the entire system (this accounts
for the terms linear in $\beta$). 

The second term in Eq.~(\ref{A0b}) is the Coulomb energy density,
which, since we are neglecting Coulomb exchange (see Section IV), is given
simply by
\beqy\label{A1a}
\mathcal{E}_{\rm Coul} = \frac{1}{2}e\rho_{\rm ch} V^{\rm Coul} \, ,
\eeqy
in which $e\rho_{\rm ch}$ is the charge density associated with protons
(this differs from $e\rho_p$ because we are taking account of the finite
size of the proton \cite{cgp08}), and $V^{\rm Coul}$ is the electrostatic 
potential, given by
\beqy\label{A1b}
V^{\rm Coul}(\pmb{r}) = e \int d^3\pmb{r^\prime}\,
\frac{\rho_{\rm ch}(\pmb{r^\prime})}{|\pmb{r}-\pmb{r^\prime}|} \, .
\eeqy
The last term in Eq.~(\ref{A0b}) is the pairing-energy density,
discussed fully in Refs.~\cite{cgp08,gcp09}.

{\it Self-consistent s.p. fields.}
In coordinate-space (assuming time-reversal invariance), the HFB equations read
\beqy\label{B1}
\sum_{\sigma^\prime=\pm1}
\begin{pmatrix} h^\prime_q(\pmb{r} )_{\sigma \sigma^\prime} -\lambda_q\, \delta_{\sigma \sigma^\prime} & \Delta_q(\pmb{r}) \delta_{\sigma \sigma^\prime} \\ \Delta_q(\pmb{r}) \delta_{\sigma \sigma^\prime} & -h^\prime_q(\pmb{r})_{\sigma \sigma^\prime} 
+ \lambda_q\, \delta_{\sigma \sigma^\prime} \end{pmatrix}\begin{pmatrix} 
\psi^{(q)}_{1i}(\pmb{r},\sigma^\prime) \\ \psi^{(q)}_{2i}(\pmb{r},\sigma^\prime) \end{pmatrix} =
E_i \begin{pmatrix} \psi^{(q)}_{1i}(\pmb{r},\sigma) \\ \psi^{(q)}_{2i}(\pmb{r},\sigma) \end{pmatrix} \, ,
\eeqy
where the s.p. Hamiltonian $h^\prime_q(\pmb{r} )_{\sigma \sigma^\prime}$ and 
pairing field $\Delta_q(\pmb{r})$  are given by 
\beqy\label{B2}
h^\prime_q(\pmb{r})_{\sigma^\prime\sigma} \equiv -\bfdel\cdot
\frac{\hbar^2}{2M_q^*(\pmb{r})}\bfdel\, \delta_{\sigma\sigma^\prime}
+ U_q(\pmb{r}) \delta_{\sigma\sigma^\prime}
-{\rm i}\pmb{W_q}(\pmb{r}) \cdot\bfdel\times\mbox{\boldmath$\sigma$}_{\sigma^\prime\sigma}
\eeqy
and
\beqy\label{B3}
\Delta_q(\pmb{r})=\frac{\partial \mathcal{E}_{\rm HFB}(\pmb{r})}{\partial \tilde{\rho}_q(\pmb{r})}=\frac{1}{2}v^{\pi q} [\rho_n(\pmb{r}),\rho_p(\pmb{r})] \tilde{\rho}_q(\pmb{r}) \, .
\eeqy

The s.p. fields appearing in Eq.~(\ref{B2}) are defined by
\beqy\label{B4}
\frac{\hbar^2}{2M_q^*(\pmb{r})} =
\frac{\partial \mathcal{E}_{\rm HFB}(\pmb{r})}{\partial\tau_q(\pmb{r})}\, ,
\hskip0.5cm
U_q(\pmb{r})=\frac{\partial \mathcal{E}_{\rm HFB}(\pmb{r})}{\partial\rho_q(\pmb{r})} - \pmb{\nabla}\cdot\frac{\partial \mathcal{E}_{\rm HFB}(\pmb{r})}{\partial (\pmb{\nabla}\rho_q(\pmb{r}))}\, , \hskip0.5cm
\pmb{W}_q(\pmb{r})=\frac{\partial \mathcal{E}_{\rm HFB}(\pmb{r})}
{\partial\pmb{J}_q(\pmb{r})}  \, .
\eeqy
From Eq.~(\ref{A1}), we find
\beqy
\label{B5}
\frac{\hbar^2}{2M_q^*} &=&
\frac{\hbar^2}{2M_q} + \frac{1}{4} t_1 \Biggl[ \left(1+ \frac{1}{2} x_1\right) \rho 
- \left( \frac{1}{2} + x_1\right)\rho_q\Biggr]  
 + \frac{1}{4} t_2 \Biggl[ \left(1+ \frac{1}{2} x_2\right) \rho + \left( \frac{1}{2} + x_2\right)\rho_q\Biggr]\nonumber \\
&+ &\frac{1}{4} t_4 \Biggl[ \left(1+ \frac{1}{2} x_4\right) \rho 
- \left( \frac{1}{2} + x_4\right)\rho_q\Biggr] \rho^\beta
 + \frac{1}{4} t_5 \Biggl[ \left(1+ \frac{1}{2} x_5\right) \rho + \left( \frac{1}{2} + x_5\right)\rho_q\Biggr]\rho^\gamma \, , \nonumber\\ 
\eeqy

\beqy\label{B6}
U_q&=& t_0\Biggl[\left(1+ \frac{1}{2} x_0\right)\rho - \left(\frac{1}{2} +
x_0\right)\rho_q\Biggr] \nonumber \\
&+&\frac{1}{4} t_1 \Biggl[\left(1+ \frac{1}{2} x_1\right) \left(\tau - 
\frac{3}{2}\nabla^2\rho\right) - \left(\frac{1}{2} +x_1\right)
\left(\tau_q - \frac{3}{2}\nabla^2\rho_q\right) \Biggr]  \nonumber \\
&+&\frac{1}{4} t_2 \Biggl[\left(1+ \frac{1}{2} x_2\right) 
\left(\tau + \frac{1}{2}\nabla^2\rho\right) + \left(\frac{1}{2} +x_2\right)
\left(\tau_q + \frac{1}{2}\nabla^2\rho_q\right) \Biggr]  \nonumber \\
&+&\frac{1}{12} t_3 \Biggl[\left(1+ \frac{1}{2} x_3\right) (2+\alpha)
\rho^{\alpha+1} - \left(\frac{1}{2} +x_3\right)\left(2\rho^\alpha \rho_q
+\alpha \rho^{\alpha-1} \sum_{q^\prime=n,p} \rho_{q^\prime}^2 \right) \Biggr]  
\nonumber \\
&+&\frac{1}{8} t_4\Bigg[\left(1+\frac{1}{2}x_4\right)\rho^{\beta-1}
\Big\{2(1+\beta)\rho\tau-(2\beta+3)\Big(\frac{1}{2}\beta(\nabla\rho)^2
+\rho\nabla^2\rho\Big)\Big\} \nonumber\\
&+&\left(\frac{1}{2}+x_4\right)\rho^{\beta-2}
\Big\{3\beta\rho\pmb{\nabla}\rho\cdot\pmb{\nabla}\rho_q 
+3\rho^2\nabla^2\rho_q-2\rho^2\tau_q \nonumber \\
&+&\beta(\beta-1)\rho_q(\nabla\rho)^2+\beta\rho\rho_q\nabla^2\rho
-\frac{1}{2}\beta\rho\sum_{q^\prime=n,p}\Big[(\nabla\rho_{q^\prime})^2
+4\rho_{q^\prime}\tau_{q^\prime}-2\rho_{q^\prime}\nabla^2\rho_{q^\prime}
\Big]\Big\}\Bigg]\nonumber\\
&+&\frac{1}{4}t_5\Bigg[\left(1+\frac{1}{2}x_5\right)\Big\{(1+\gamma)\rho\tau +
\frac{1}{4}\gamma(\nabla\rho)^2+\frac{1}{2}\rho\nabla^2\rho\Big\}\nonumber\\
&+&\left(\frac{1}{2}+x_5\right)\Big\{\rho\tau_q+\frac{1}{2}\rho\nabla^2\rho_q
+\gamma\sum_{q^\prime=n,p}\left\{\rho_{q^\prime}\tau_{q^\prime}
-\frac{1}{4}(\nabla\rho_{q^\prime})^2\right\}
+\frac{1}{2}\gamma\,\pmb{\nabla}\rho\cdot\pmb{\nabla}\rho_q\Big\}
\Bigg]\rho^{\gamma-1} \nonumber\\
&-& \frac{1}{16} (t_4 x_4\beta\rho^{\beta-1} + 
t_5 x_5\gamma\rho^{\gamma-1}) J^2 + \frac{1}{16} (t_4\beta\rho^{\beta-1}-
t_5\gamma\rho^{\gamma-1})\sum_{q^\prime=n,p}J_{q^\prime}^2\nonumber \\
&-&\frac{1}{2} W_0 \left(\pmb{\nabla}\cdot \pmb{J} + \pmb{\nabla}\cdot \pmb{J_q}\right)+ \delta_{q,p} V^{\rm Coul}+\frac{1}{4} \sum_{q^\prime=n,p}\frac{\partial v^{\pi q^\prime}}{\partial \rho_q}\, \tilde{\rho}_{q^\prime}^2 
\eeqy
and

\beqy\label{B7}
\pmb{W_q}&=&\frac{1}{2} W_0 \pmb{\nabla} (\rho+\rho_q) -\frac{1}{8} (t_1 x_1 + t_2 x_2) \pmb{J} + \frac{1}{8} (t_1-t_2) \pmb{J_q} \nonumber \\
&-& \frac{1}{8} (t_4 x_4 \rho^\beta + t_5 x_5 \rho^\gamma) \pmb{J} + \frac{1}{8} (t_4 \rho^\beta-t_5\rho^\gamma) \pmb{J_q}\, .
\eeqy

The Coulomb field $V^{\rm Coul}(\pmb{r})$ is given by Eq.~(\ref{A1b}).

{\it Unpolarized homogeneous nuclear matter.} 
For the energy per nucleon of infinite nuclear matter of density $\rho$ and
asymmetry $\eta$ (defined by $\eta = (\rho_n - \rho_p)/\rho)$, Eq.~(\ref{A1}) 
reduces to
\beqy
\label{C1} 
e &=& \frac{3\hbar^2}{20}k_F^2\left\{\frac{1}{M_n}(1+\eta)^{5/3}+
\frac{1}{M_p}(1-\eta)^{5/3}\right\}    \nonumber\\
&+&\frac{1}{8}t_0\Biggl[3 - (2x_0+1)\eta^2\Biggr]\rho \nonumber \\
&+&\frac{3}{40}t_1\Biggl[(2+x_1)F_{5/3}(\eta) -(\frac{1}{2}+x_1)F_{8/3}(\eta)
\Biggr]\rho\,k_F^2 \nonumber\\
&+&\frac{3}{40}t_2\Biggl[(2+x_2)F_{5/3}(\eta) +(\frac{1}{2}+x_2)F_{8/3}(\eta)
\Biggr]\rho\,k_F^2 \nonumber\\
&+&\frac{1}{48}t_3\Biggl[3-(1+2x_3)\eta^2\Biggr]\rho^{\alpha+1} \nonumber\\
&+&\frac{3}{40}t_4\Biggl[(2+x_4)F_{5/3}(\eta) - (\frac{1}{2}+x_4)F_{8/3}(\eta)
\Biggr]\rho^{\beta+1}\,k_F^2 \nonumber\\
&+&\frac{3}{40}t_5\Biggl[(2+x_5)F_{5/3}(\eta) + (\frac{1}{2}+x_5)F_{8/3}(\eta)
\Biggr]\rho^{\gamma+1}\,k_F^2 \quad , \nonumber\\
\eeqy
where
\beqy
\label{C2}
k_F= \left(\frac{3\pi^2\rho}{2}\right)^{1/3}
\eeqy
and
\beqy
\label{C3}
F_x(\eta)= \frac{1}{2}\Biggl[(1+\eta)^x+(1-\eta)^x\Biggr]\quad .
\eeqy

For SNM ($\eta$ = 0) Eq.~(\ref{C1}) reduces to
\beqy
\label{C4}
e(\eta = 0) &=& \frac{3\hbar^2}{10M}k_F^2+\frac{3}{8}t_0\rho
+\frac{3}{80}\Biggl[3t_1+t_2(5+4x_2)\Biggr]\rho\,k_F^2
+\frac{1}{16}t_3\rho^{\alpha+1} \nonumber\\
&+&\frac{9}{80}t_4\rho^{\beta+1}\,k_F^2 
+\frac{3}{80}t_5(5+4x_5)\rho^{\gamma+1}\,k_F^2  \quad ,
\eeqy
where
\beqy
\label{C4A}
\frac{2}{M} = \frac{1}{M_n} + \frac{1}{M_p}  \quad  .
\eeqy
From Eq.~(\ref{C4A})we have
\beqy
\label{C5}
k_F^2\frac{\partial^2 e }{\partial k_F^2}\Biggr\vert_{\eta=0}&=&\frac{3\hbar^2}{5M}k_F^2
+\frac{9}{4}t_0\rho 
+\frac{3}{4}\Biggl[3t_1+t_2(5+4x_2)\Biggr]\rho\,k_F^2 \nonumber\\
&+&\frac{3}{16}t_3(\alpha+1)(3\alpha+2)\rho^{\alpha+1} \nonumber\\
&+&\frac{9}{80}t_4(3\beta+5)(3\beta+4)\rho^{\beta+1}k_F^2 \nonumber\\
&+&\frac{3}{80}t_5(5+4x_5)(3\gamma+5)(3\gamma+4)\rho^{\gamma+1}k_F^2 
\quad . \nonumber\\
\eeqy
Evaluating this at $\rho = \rho_0$ and introducing $k_{F0}= \left(3\pi^2\rho_0/2\right)^{1/3}$
defines the usual incompressibility
\beqy
\label{C6}
K_v = k_{F0}^2\frac{\partial^2 e}{\partial k_F^2}\Biggr\vert_{\eta=0,\,\rho=\rho_0}
= 9\rho_0^2\frac{\partial^2 e}{\partial \rho_0^2}\Biggr\vert_{\eta=0,\,\rho=\rho_0} \quad ,
\eeqy
the two forms being equivalent only because at $\rho = \rho_0$ we have
\beqy
\label{C7}
\frac{\partial e }{\partial\rho}\Biggr\vert_{\eta=0,\,\rho=\rho_0} = 0 \quad .
\eeqy

Expanding $(1 \pm \eta)^x$ up to $\eta^2$ allows us to rewrite Eq.~(\ref{C1}) 
as
\beqy
\label{C8}
e=e(\eta=0)+\frac{\hbar^2}{4}k_F^2\left(\frac{1}{M_n} - 
\frac{1}{M_p}\right)\eta + e_{sym}\eta^2+O\Big(\eta^4\Big) \quad ,
\eeqy 
where we have introduced the symmetry energy
\beqy
\label{C9}
e_{sym}(\rho) &=& \frac{\hbar^2}{6M}k_F^2-\frac{1}{8}t_0(2x_0+1)\rho 
+\frac{1}{24}\Biggl[-3t_1x_1+t_2(4+5x_2)\Biggr]\rho\,k_F^2 \nonumber \\
&-&\frac{1}{48}t_3(1+2x_3)\rho^{\alpha+1} 
-\frac{1}{8}t_4x_4\rho^{\beta+1}\,k_F^2
+\frac{1}{24}t_5(4+5x_5)\rho^{\gamma+1}\,k_F^2  
\eeqy
(note the breaking of charge symmetry implied by the neutron-proton mass
difference; this is explicitly included in the finite-nucleus calculations).
The usual symmetry coefficient \cite{ms69} is then given by 
\beqy
\label{C10}
J = e_{sym}(\rho_0) \quad .
\eeqy
For the density-symmetry coefficient \cite{ms69} defined by
\beqy
\label{C11}
L = k_{F0}\frac{{\rm d}e_{sym}}{{\rm d}k_F}\Biggr\vert_{\rho=\rho_0}
=3\rho_0\frac{{\rm d}e_{sym}}{{\rm d}\rho}\Biggr\vert_{\rho=\rho_0} \quad ,
\eeqy
it follows from Eq.~(\ref{C9}) that
\beqy
\label{C12}
L &=& \frac{\hbar^2}{3M}k_{F0}^2-\frac{3}{8}t_0(2x_0+1)\rho_0
+\frac{5}{24}\Biggl[-3t_1x_1+t_2(4+5x_2)\Biggr]\rho_0\,k_{F0}^2 \nonumber \\
&-&\frac{\alpha+1}{16}t_3(1+2x_3)\rho_0^{\alpha+1}
-\frac{5+3\beta}{8}t_4x_4\rho_0^{\beta+1}\,k_{F0}^2  \nonumber \\
&+&\frac{5+3\gamma}{24}t_5(4+5x_5)\rho_0^{\gamma+1}\,k_{F0}^2 \quad .
\eeqy

Setting $\eta$ = 1 in Eq.~(\ref{C1}) gives us for the energy per nucleon in 
unpolarized NeuM
\beqy
\label{C13}
e &=& \frac{3\hbar^2}{10M_n}k_{Fn}^2+\frac{1}{4}t_0(1-x_0)\rho
+\frac{3}{40}t_1(1-x_1)\rho\,k_{Fn}^2
+\frac{9}{40}t_2(1+x_2)\rho\,k_{Fn}^2 \nonumber\\
&+&\frac{1}{24}t_3(1-x_3)\rho^{\alpha+1}
+\frac{3}{40}t_4(1-x_4)\rho^{\beta+1}\,k_{Fn}^2
+\frac{9}{40}t_5(1+x_5)\rho^{\gamma+1}\,k_{Fn}^2 \quad , 
\eeqy
where 
\beqy
\label{C14}
k_{Fn}= (3\pi^2\rho)^{1/3} 
\eeqy
(Eq.~(\ref{C13}) cannot be derived from Eqs.~(\ref{C8}) and (\ref{C9})).

Note that the terms in $t_4$ and $t_5$ will not contribute to any of the
foregoing expressions for homogeneous unpolarized matter if the conditions (\ref{9}), 
(\ref{11}) and (\ref{12}) are satisfied.

{\it Polarized nuclear matter}

In general the Skyrme energy density can be decomposed into a 
time-even part $\mathcal{E}_{\rm Sky}^{\rm even}$ given by 
Eq.~(\ref{A1}) and a time-odd part $\mathcal{E}_{\rm Sky}^{\rm odd}$ 
which is non-zero only when time-reversal invariance is not satisfied.
In polarized homogeneous nuclear matter these two parts take the form
\beqy\label{D1}
\mathcal{E}_{\rm Sky}^{\rm even}=\sum_{q=n,p}\frac{\hbar^2}{2M_q}\tau_q+C_0^\rho \rho^2+C_1^\rho (\rho_n-\rho_p)^2+C_0^\tau \rho\tau+C_1^\tau (\rho_n-\rho_p)(\tau_n-\tau_p)\quad ,
\eeqy
and
\beqy\label{D2}
\mathcal{E}_{\rm Sky}^{\rm odd}=C_0^s \pmb{s}^2+C_1^s(\pmb{s_n}-\pmb{s_p})^2+C_0^T\pmb{s}\cdot\pmb{T}
+C_1^T (\pmb{s_n}-\pmb{s_p})\cdot(\pmb{T_n}-\pmb{T_p})
\eeqy
where $\pmb{s_q}$ and $\pmb{T_q}$ are the spin density and kinetic spin density respectively
(for a precise definition, see, for example, 
Bender \textit{et al}.~\cite{bend02}), and 
$\pmb{s}=\pmb{s_n}+\pmb{s_p}$, $\pmb{T}=\pmb{T_n}+\pmb{T_p}$. 

The expressions for the coefficients appearing in Eqs.~(\ref{D1}) and 
(\ref{D2}) in the case of the conventional Skyrme force~(\ref{1}) can be found,
for example, in Appendix B of Ref.~\cite{bend02}. The coefficients $C_0^\rho$, 
$C_1^\rho$, $C_0^s$ and $C_1^s$ depend only on the $t_0$ and $t_3$ terms 
of the Skyrme force~(\ref{1}) and therefore remain unchanged when the new terms
of Eq.~(\ref{2}) are included. This is not the case for the other coefficients.
However $C_0^\tau$ and $C_1^\tau$ can be readily obtained by comparing 
Eqs.~(\ref{D1}) and (\ref{A1}). The expressions for the remaining coefficients 
$C_0^T$ and $C_1^T$ 
can also be obtained from Eq.~(\ref{A1}) using the gauge invariance of the 
Skyrme force~\cite{eng75,doba95}: $-C_0^T$ and $-C_1^T$ 
coincide with the coefficients in front of the terms proportional to $J^2$ and $(\pmb{J_n}-\pmb{J_p})^2$ in Eq.~(\ref{A1}). 
The various coefficients are thus given by
\bmlet
\beqy\label{A30a}
C_0^\rho=\frac{3}{8}t_0+\frac{3}{48}t_3\rho^\alpha
\eeqy

\beqy
C_1^\rho=-\frac{1}{4}t_0\left(\frac{1}{2}+x_0\right)-\frac{1}{24}t_3(1+x_3)\rho^\alpha
\eeqy

\beqy
C_0^s=-\frac{1}{4}t_0\left(\frac{1}{2}-x_0\right)-\frac{1}{24}t_3\left(\frac{1}{2}-x_3\right)\rho^\alpha
\eeqy

\beqy
C_1^s=-\frac{1}{8}t_0-\frac{1}{48}t_3\rho^\alpha
\eeqy

\beqy
C_0^\tau=\frac{3}{16}t_1+\frac{1}{4}t_2\left(\frac{5}{4}+x_2\right)+\frac{3}{16}t_4 \rho^\beta +\frac{1}{4}t_5\left(\frac{5}{4}+x_5\right)\rho^\gamma
\eeqy

\beqy
C_1^\tau=-\frac{1}{8}t_1\left(\frac{1}{2}+x_1\right)+\frac{1}{8}t_2\left(\frac{1}{2}+x_2\right)-\frac{1}{8}t_4 \rho^\beta \left(\frac{1}{2}+x_4\right)+\frac{1}{8}t_5\rho^\gamma\left(\frac{1}{2}+x_5\right)
\eeqy

\beqy
C_0^T=-\frac{1}{8}\biggl[t_1\left(\frac{1}{2}-x_1\right)-t_2\left(\frac{1}{2}+x_2\right)+t_4\rho^\beta\left(\frac{1}{2}-x_4\right)-t_5\rho^\gamma\left(\frac{1}{2}+x_5\right)\biggr]
\eeqy

\beqy\label{A30h}
C_1^T=-\frac{1}{16}(t_1-t_2) -\frac{1}{16}(t_4 \rho^\beta-t_5\rho^\gamma)\, . 
\eeqy
\emlet
Under the constraints~(\ref{9}), (\ref{10}) and (\ref{11}), the new terms in 
the Skyrme force do not affect the coefficients $C_0^\tau$ and $C_1^\tau$. In 
the particular case of model HFB-18 for which $t_4=t_5$, the coefficient 
$C_1^T$ is also unchanged, so that only $C_0^T$ changes. 

It can thus be seen that the expressions of the above $C$-coefficients given in
Ref.\cite{bend02} can be extended to the generalized Skyrme force~(\ref{5}) 
simply by making the following substitutions: 
\bmlet
\beqy\label{D3a}
t_1 \rightarrow t_1 + t_4\rho^\beta \quad ,
\eeqy
\beqy\label{D3b}
t_1 x_1 \rightarrow t_1 x_1 + t_4 x_4\rho^\beta \quad ,
\eeqy
\beqy\label{D3c}
t_2 \rightarrow t_2 + t_5\rho^\gamma \quad ,
\eeqy
\beqy\label{D3d}
t_2 x_2 \rightarrow t_2 x_2 + t_5 x_5\rho^\gamma \quad .
\eeqy
\emlet
With this simple rule, the expression for the energy per nucleon of 
asymmetric polarized homogeneous nuclear matter can be easily obtained
from Eq.~(C.14) of Ref.~\cite{bend02}.

{\it Landau parameters}

The dimensionless Landau parameters for symmetric nuclear matter corresponding 
to the generalized Skyrme force~(\ref{5}) are given in terms of the 
C-coefficients of Eqs.~(\ref{A30a}) - (\ref{A30h})by 
\bmlet
\beqy
F_0=N_0\Biggl[2 C_0^\rho+2 C_0^\tau k_{\rm F}^2+4\rho \frac{{\rm d} C_0^\rho}{{\rm d}\rho}+\rho^2\frac{{\rm d}^2 C_0^\rho}{{\rm d}\rho^2}+\rho\tau\frac{{\rm d}^2 C_0^\tau}{{\rm d}\rho^2}+\frac{{\rm d} C_0^\tau}{{\rm d}\rho}(2\tau+2\rho k_{\rm F}^2)\Biggr]
\eeqy
\beqy
F_0^\prime=N_0\Biggl[2 C_1^\rho+2 C_1^\tau k_{\rm F}^2\Biggr]
\eeqy
\beqy
F_1=-2 N_0 C_0^\tau k_{\rm F}^2
\eeqy
\beqy
F_1^\prime=-2 N_0 C_1^\tau k_{\rm F}^2
\eeqy
\beqy
G_0=N_0\Biggl[2 C_0^s+2 C_0^T k_{\rm F}^2\Biggr]
\eeqy
\beqy
G_0^\prime=N_0\Biggl[2 C_1^s+2 C_1^T k_{\rm F}^2\Biggr]
\eeqy
\beqy
G_1=-2 N_0 C_0^T k_{\rm F}^2
\eeqy
\beqy
G_1^\prime=-2 N_0 C_1^T k_{\rm F}^2
\eeqy
\emlet
where $N_0$ is the density of s.p. states at the Fermi level
\beqy
N_0=\frac{2 M^*_s k_{\rm F}}{\hbar^2\pi^2}\, .
\eeqy
For the conventional Skyrme force~(\ref{1}), the above expressions reduce to 
those given in Appendix D of Ref.\cite{bend02}. Since with BSk18 $C_0^T$ is the
only C-coefficient that is modified by the $t_4$ and $t_5$ terms it follows 
that they affect only the Landau parameters $G_0$ and $G_1$.

\newpage

\begin{table}
\centering
\caption{Force BSk18: lines 1-16 show the Skyrme parameters, lines 17-21
the pairing parameters, and the last four lines the Wigner parameters
(see text for further details). For convenience of comparison we also show
model HFB-17 \cite{gcp09} (note that in the latter reference there were 
misprints in the pairing parameters $f_q^{\pm}$).} 
\label{tab1}
\vspace{.5cm}
\begin{tabular}{|c|cc|}
\hline
  &HFB-18&HFB-17\\
\hline
  $t_0$ {\scriptsize [MeV fm$^3$]}   & -1837.96 & -1837.33 \\
  $t_1$ {\scriptsize [MeV fm$^5$]}   & 428.880 &389.102  \\
  $t_2$ {\scriptsize [MeV fm$^5$]}   & -3.23704 &-3.17417\\
  $t_3$ {\scriptsize [MeV fm$^{3+3\alpha}$]}  & 11528.9 & 11523.8 \\
  $t_4$ {\scriptsize [MeV fm$^{5+3\beta}$]}  & -400.000 &- \\
  $t_5$ {\scriptsize [MeV fm$^{5+3\gamma}$]}  & -400.000 &- \\
  $x_0$                              &  0.421290&0.411377  \\
  $x_1$                              & -0.907175&-0.832102\\
  $x_2$                              & 57.7185 &49.4875 \\
  $x_3$                              &  0.683926 &0.654962 \\
  $x_4$                              & -2.00000  &-\\
  $x_5$                              &  -2.00000 &-\\
  $W_0$ {\scriptsize [MeV fm$^5$]}   &  138.904  &145.885  \\
  $\alpha$                           &  0.3 &0.3  \\
  $\beta$                            &1.0 &- \\
  $\gamma$                           &1.0  &-\\
  $f_{n}^+$  &  1.00 & 1.00\\
  $f_{n}^-$  &  1.06& 1.05  \\
  $f_{p}^+$  &  1.04 &1.04 \\
  $f_{p}^-$  &  1.09 &1.10  \\
  $\varepsilon_{\Lambda}$ {\scriptsize [MeV]}  &  16.0 &16.0   \\
  $V_W$ {\scriptsize [MeV]}           & -2.10&-2.00   \\
  $\lambda$                           & 340 &320   \\
  $V_W^{\prime}$ {\scriptsize [MeV]}  & 0.74&0.86   \\
  $A_0$                               &28 &28  \\
 \hline
\end{tabular}
\end{table}

\begin{table}
\centering
\caption{Parameters of Eq.(\ref{15}) for collective correction to model HFB-18.}
\label{tab2}
\vspace{.5cm}
\begin{tabular}{|c|c|}
\hline
$b$  (MeV)&  0.8\\
$c$ & 10.0 \\
$d$ (MeV) & 3.0 \\
$l$ & 16.0 \\
$\beta_2^0$ & 0.1\\
 \hline
\end{tabular}
\end{table}

\begin{table}
\centering
\caption{Rms ($\sigma$) and mean ($\bar{\epsilon}$) deviations between data and
predictions for model HFB-18; for convenience of comparison we also show 
model HFB-17 \cite{gcp09}. The first pair of lines refers to all the 2149 
measured masses $M$ that were fitted \cite{audi03}, the second pair to the 
masses 
$M_{nr}$ of the subset of 185 neutron-rich nuclei with $S_n \le $ 5.0 MeV, the 
third pair to the neutron separation energies $S_n$ (1988 measured values), the
fourth pair to beta-decay energies $Q_\beta$  (1868 measured values) and
the fifth pair to charge radii (782 measured values \cite{ang04}). 
The last line shows
the calculated neutron-skin thickness of $^{208}$Pb for these models.}
\label{tab3}
\vspace{.5cm}
\begin{tabular}{|c|cc|}
\hline
&HFB-18&HFB-17  \\
\hline
$\sigma(M)$ {\scriptsize [MeV]} &0.585 &0.581\\
$\bar{\epsilon}(M)$ {\scriptsize [MeV]}&0.007 &-0.019\\
$\sigma(M_{nr})$ {\scriptsize [MeV]}&0.758 &0.729\\
$\bar{\epsilon}(M_{nr})$ {\scriptsize [MeV]}&0.172&0.119\\
$\sigma(S_n)$ {\scriptsize [MeV]}&0.487&0.506\\
$\bar{\epsilon}(S_n)$ {\scriptsize [MeV]}&-0.012&-0.010\\
$\sigma(Q_\beta)$ {\scriptsize [MeV]}&0.561   &0.583\\
$\bar{\epsilon}(Q_\beta)$ {\scriptsize [MeV]}&0.025&0.022\\
$\sigma(R_c)$ {\scriptsize [fm]}&0.0274&0.0300\\
$\bar{\epsilon}(R_c)$ {\scriptsize [fm]}&0.0016&-0.0114\\
$\theta$($^{208}$Pb) {\scriptsize [fm]}&0.15&0.15\\
\hline
\end{tabular}
\end{table}

\begin{table}
\centering
\caption{Parameters of infinite nuclear matter for force BSk18; for 
convenience of comparison we also show force BSk17 \cite{gcp09}.} 
\label{tab4}
\vspace{.5cm}
\begin{tabular}{|c|cc|}
\hline
&BSk18&BSk17\\
\hline
$a_v$ {\scriptsize [MeV]}&-16.063&-16.054 \\
$\rho_0$ {\scriptsize [fm$^{-3}$]}&0.1586&0.1586 \\
$J$ {\scriptsize [MeV]}&30.0&30.0  \\
$K_v$ \scriptsize [MeV]&241.8& 241.7 \\
$L$ \scriptsize [MeV]&36.21 &36.28  \\
$M^*_s/M$ &0.80& 0.80 \\
$M^*_v/M$&0.79& 0.78     \\
$F_0$ &-0.12&-0.12\\
$F_0^{'}$&0.97&0.97\\
$F_1$&-0.60&-0.60\\
$F1^{'}$ &0.032&0.068\\
$G_0$ &-0.33&-0.69\\
$G_0^{'}$&0.46&0.50\\
$G_1$&1.23&1.55\\
$G_1^{'}$ &0.50&0.45\\
\hline
\end{tabular}
\end{table}
\eject

\newpage

\end{document}